\documentstyle[12pt,ssi]{article}
\def\Re{{\cal R \mskip-4mu \lower.1ex \hbox{\it e}\,}}
\def\im{{\cal I \mskip-5mu \lower.1ex \hbox{\it m}\,}}
\def\ie{{\it i.e.}}

\def\etal{{\it et al.}}

\def\tev{\,{\rm TeV}}
\def\gev{\,{\rm GeV}}

\def\to{\rightarrow}

\def\mh{\ifmmode m\sbl H \else $m\sbl H$\fi}
\def\mch{\ifmmode m_{H^\pm} \else $m_{H^\pm}$\fi}
\def\mt{\ifmmode m_t\else $m_t$\fi}
\def\mc{\ifmmode m_c\else $m_c$\fi}
\def\mz{\ifmmode M_Z\else $M_Z$\fi}
\def\mw{\ifmmode M_W\else $M_W$\fi}
\def\mws{\ifmmode M_W^2 \else $M_W^2$\fi}
\def\mhs{\ifmmode m_H^2 \else $m_H^2$\fi}
\def\mzs{\ifmmode M_Z^2 \else $M_Z^2$\fi}
\def\mts{\ifmmode m_t^2 \else $m_t^2$\fi}
\def\mcs{\ifmmode m_c^2 \else $m_c^2$\fi}
\def\mchs{\ifmmode m_{H^\pm}^2 \else $m_{H^\pm}^2$\fi}
\def\ztwo{\ifmmode Z_2\else $Z_2$\fi}
\def\zone{\ifmmode Z_1\else $Z_1$\fi}
\def\mtwo{\ifmmode M_2\else $M_2$\fi}
\def\mone{\ifmmode M_1\else $M_1$\fi}
\def\tb{\ifmmode \tan\beta \else $\tan\beta$\fi}
\def\xw{\ifmmode x\sub w\else $x\sub w$\fi}
\def\ch{\ifmmode H^\pm \else $H^\pm$\fi}
\def\bsg{\ifmmode b\to s\gamma\else $b\to s\gamma$\fi}
\def\bkg{\ifmmode B\to K^*\gamma\else $B\to K^*\gamma$\fi}
\def\lum{\ifmmode {\cal L}\else ${\cal L}$\fi}
\def\inpb{\ifmmode {\rm pb}^{-1}\else ${\rm pb}^{-1}$\fi}
\def\infb{\ifmmode {\rm fb}^{-1}\else ${\rm fb}^{-1}$\fi}
\def\epem{\ifmmode e^+e^-\else $e^+e^-$\fi}
\def\ppb{\ifmmode \bar pp\else $\bar pp$\fi}

\newskip\zatskip \zatskip=0pt plus0pt minus0pt
\def\matth{\mathsurround=0pt}
\def\lsim{\mathrel{\mathpalette\atversim<}}
\def\gsim{\mathrel{\mathpalette\atversim>}}
\def\atversim#1#2{\lower0.7ex\vbox{\baselineskip\zatskip\lineskip\zatskip
  \lineskiplimit 0pt\ialign{$\matth#1\hfil##\hfil$\crcr#2\crcr\sim\crcr}}}

\begin{document}
\rightline{\vbox{\halign{&#\hfil\cr
&SLAC-PUB-6521\cr
&May 1994\cr
&T/E\cr}}}

\title{Top Ten Models Constrained by \bsg }

\author{JoAnne L. Hewett\thanks{Work supported by the Department of
Energy, Contract DE-AC03-76SF00515.}\\
Stanford Linear Accelerator Center\\
Stanford University, Stanford, CA   94309\\
}

\maketitle

\begin{abstract}

The radiative decay \bsg\ is examined in the Standard Model and in nine
classes of models which contain physics beyond the Standard Model.  The
constraints which may be placed on these models from the recent results
of the CLEO Collaboration on both inclusive and exclusive radiative $B$
decays is summarized.  Reasonable bounds are found for the parameters
in some cases.

\vskip1.25in

\noindent{Presented at the {\it 21st Annual SLAC Summer Institute on Particle
Physics:  Spin Structure in High Energy Processess} (School: 26 Jul - 3 Aug,
Topical Conference: 4-6 Aug), Stanford, CA, 26 Jul - 6 Aug 1993.}

\end{abstract}

\section{Introduction}

The Standard Model (SM) of electroweak interactions is in complete
agreement with present experimental data\cite{lep}.  Nonetheless, it is
believed to leave many questions unanswered, and this belief has resulted in
numerous attempts to discover a more fundamental underlying theory.  The
search for new physics is conducted via a three-prong attack: (i) direct
production of new particles at high energy colliders, (ii) deviations from
SM predictions in precision measurements, and (iii) indirect observation of
new physics in rare or forbidden processes.
The first approach relies on a discovery via the direct production
of exotic particles or observation of new reactions.  The second
and third techniques offer a complementary strategy by searching for
indirect effects of new physics in higher order processes.  In particular,
the probing of loop induced couplings can provide a means of testing the
detailed structure of the SM at the level of radiative corrections where the
Glashow-Iliopoulos-Maiani\cite{gim} (GIM) cancellations are important.
This talk will focus on the latter option, and will examine the
radiative decay \bsg.

Radiative $B$ decays are one of the best testing grounds of the SM due to
recent progress on both theoretical and experimental fronts.  The
CLEO Collaboration has
observed\cite{cleo}  the exclusive decay \bkg\ with a branching
fraction of $B(\bkg )=(4.5\pm 1.5\pm 0.9)\times 10^{-5}$ and has also placed
an upper limit on the underlying quark-level process of
$B(\bsg )<5.4\times 10^{-4}$ at the $95\%$ C.L. Using a conservative value of
the ratio of exclusive to inclusive decay rates based
on lattice calculations\cite{soni},  the observation of the
exclusive process also implies the lower bound $B(\bsg )>0.65\times 10^{-4}$
at $95\%$ C.L.  On the theoretical side, the reliability of the calculation
of the quark-level process \bsg\ is improving as partial calculations of
the next-to-leading logarithmic QCD corrections to the
effective Hamiltonian now exist.\cite{misiak}
These new results have inspired a large number of
investigations of this decay in various classes of models, which can be
summarized by the following list:
\begin{itemize}
\item ``Top Ten'' Models Constrained by \bsg
\begin{tabbing}
3. Anomalous Trilinear Gauge Couplings \= 8. Extended Technicolor \kill
1. Standard Model  \> 6. Supersymmetry\\
2. Anomalous Top-Quark Couplings  \> 7. Three-Higgs-Doublet Model\\
3. Anomalous Trilinear Gauge Couplings \>  8. Extended Technicolor\\
4. Fourth Generation \>  9. Leptoquarks\\
5. Two-Higgs-Doublet Models \>  10. Left-Right Symmetric Models
\end{tabbing}
\end{itemize}
In what follows, I will summarize the contributions that \bsg\ receives in
each of these models and the constraints placed on the model parameters by
the CLEO data.

\section{Models}

\subsection{Standard Model}

In the SM, the quark-level transition \bsg\ is mediated by $W$-boson and
t-quark exchange in an electromagnetic penguin diagram.  The matrix element
for this process at the electroweak scale is governed by the $\sigma_{\mu\nu}
q^\nu(1+\gamma_5)$ dipole operator.  The QCD corrections to this process are
calculated\cite{qcd} via an operator product expansion based on the effective
Hamiltonian
\begin{equation}
H_{eff}=-{4G_F\over\sqrt 2}V^*_{ts}V_{tb}\sum_{i=1}^8c_i(\mu)O_i(\mu)\,,
\end{equation}
which is then evolved from the electroweak scale down to $\mu=m_b$ by the
Renormalization Group Equations.  Here, $V_{ij}$ represents the relevant
Cabibbo-Kobayashi-Maskawa (CKM) factors.  The $O_i$ are a complete set
of renormalized dimension six operators involving light fields which govern
$b\to s$ transitions.  They consist of six four-quark operators, $O_{1-6}$,
the electromagnetic dipole operator, $O_7$, and the chromo-magnetic dipole
operator, $O_8$.  The Wilson coefficients, $c_i$, of the $b\to s$ operators
are evaluated perturbatively at the $W$ scale where the matching conditions
are imposed and are evolved down to the renormalization scale $\mu$.  The
explicit expressions for $c_{7,8}(M_W)=G_{7,8}(m_t^2/M_W^2)$ can be found in
the literature.\cite{inami}  The partial decay width is given by
\begin{equation}
\Gamma(\bsg)={\alpha G_F^2m_b^5\over 128\pi^4}|V^*_{ts}V_{tb}c_7(m_b)|^2 \,.
\end{equation}
To obtain
the branching fraction, the inclusive rate is scaled to that of the
semi-leptonic decay $b\to X\ell\nu$.  This procedure removes uncertainties
in the calculation due to an overall factor of $m_b^5$ which appears in
both expressions, and reduces the ambiguities involved with the imprecisely
determined CKM factors.  The result is then
rescaled by the experimental value\cite{drell} of
$B(b\to X\ell\nu)=0.108$.  The semi-leptonic rate is calculated incorporating
both charm and non-charm modes, and includes both phase space and QCD
corrections\cite{nicolo}.  The calculation of $c_7(m_b)$ employs the partial
next-to-leading log evolution equations from Ref.\ 5 for the
coefficients of the $b\to s$ transition operators in the effective Hamiltonian,
the ${\cal O}(\alpha_s)$ corrections due to gluon
bremsstrahlung\cite{ali}, corrections\cite{cho} for
$m_{t}>M_W$, a running $\alpha_{QED}$ evaluated at $m_b$, and the
3-loop evolution of the running $\alpha_s$ which is fitted to the global
value\cite{lep} at the $Z$ mass scale.  The ratio of CKM elements
in the scaled decay rate, $|V_{tb}V_{ts}/V_{cb}|$, is taken to be unity.

The prediction for the \bsg\ branching fraction as a function of the
top-quark  mass in the SM is shown in
Fig. 1a, taking $\mu=m_b=5\gev$.  The solid curve represents the rate with
the inclusion of the partial next-to-leading
log evolution of the operator coefficients,
while the dashed curve corresponds to the leading log case.  The effect of the
known next-to-leading order terms is to
decrease the QCD enhancements of the rate by $\sim 15\%$.
Figure 1b displays the dependency of the branching fraction (for $m_t=165\gev$)
on the choice of the renormalization scale for the Wilson coefficients.  The
uncertainty introduced by the choice of the value of $m_c/m_b$ in calculating
$B(b\to X\ell\nu)$ is also shown in this figure, where the region between
the curves corresponds to $m_c/m_b=0.316\pm 0.013$.
We see that the \bsg\ branching fraction increases by $\sim
20\%$ as the renormalization scale $\mu$ is varied from $m_b$ to $m_b/2$.
The overall variation in the SM prediction for $B(\bsg)$ due to the combined
freedom of choice in $\mu$ and $m_c/m_b$ can be as large as $30-40\%$!  Once
the
full next-to-leading order corrections have been computed, this large
dependence
on the renormalization scale will diminish.  For now, this dependence
represents an additional theoretical uncertainty\cite{bsgqcd}.  When
determining constraints on new physics from this decay, we choose
values for these parameters which yields the most conservative SM rate;  for
most of the models discussed here $\mu$ is taken to be 5.0 GeV.
Most of the parameter constraints presented here are not very sensitive to the
remaining uncertainties in the calculation of the branching
fraction arising from higher order QCD corrections, as $B(\bsg)$
is a steep function of the parameters in these cases.

\subsection{Anomalous Top-Quark Couplings}
The possibility of anomalous couplings between the top-quark and the gauge
boson
sector has been examined in the literature\cite{anomtop}.  Future colliders
such as the LHC and NLC can probe these effective couplings down to the level
of
$10^{-18}-10^{-19}$ e-cm, but they rely on direct production of top-quark
pairs, whereas \bsg\ provides the opportunity to probe the properties of
the top-quark before it is produced directly.  If the t-quark has large
anomalous couplings to on-shell photons and gluons, the resulting
prediction\cite{jlhtgr} for
the \bsg\ rate would conflict with experiment.  The most general form of
the Lagrangian which describes the interaction between top-quarks and
on-shell photons (assuming operators of dimension-five or less, only) is
\begin{equation}
{\cal L}_{t\bar t\gamma}=
e\bar t\left[ Q_t\gamma_\mu+{1\over 2m_t}\sigma_{\mu\nu}
(\kappa_\gamma+i\tilde\kappa_\gamma \gamma_5)q^\nu\right] tA^\mu + h.c. \,,
\end{equation}
where $Q_t$ is the electric charge of the t-quark, and $\kappa_\gamma
(\tilde\kappa_\gamma)$ represents the anomalous magnetic (electric)
dipole moment.  A similar expression is obtained for ${\cal L}_{t\bar tg}$.
Note that a non-vanishing value for $\tilde\kappa_\gamma$ would signal the
presence of a CP-violating amplitude.  In practice, only the coefficients
of the magnetic dipole and chromo-magnetic dipole
$b\to s$ transition operators, $O_7$ and $O_8$
respectively, are modified by the presence of these couplings.  The
coefficients of these operators at the $W$ scale can be written as
\begin{eqnarray}
c_7(M_W)& = & G_7^{SM}(\mts/\mws)+\kappa_\gamma G_1(\mts/\mws)
+i\tilde\kappa_\gamma G_2(\mts/\mws) \,, \\
c_8(M_W)& = & G_8^{SM}(\mts/\mws)+\kappa_g G_1(\mts/\mws)
+i\tilde\kappa_g G_2(\mts/\mws) \,. \nonumber
\end{eqnarray}
The functions $G_i$ are obtained by inserting the above couplings into the
Feymann diagrams in which the photon is emitted from the top-quark
line, and extracting the pure dipole-like terms after performing the loop
integrations and are given in Ref.\ 14.
All other Lorentz structures vanish due to electromagnetic
gauge invariance and the fact that the photon is on-shell.  When the
resulting branching fraction and the CLEO data are combined, the constraints
shown in Fig. 2 are obtained.  In Fig. 2a, the $95\%$ C.L.
allowed region of the anomalous magnetic
dipole operator as a function of $m_t$ lies between the curves for the cases
$\kappa_g=0$ (solid curves) and $\kappa_g=\kappa_\gamma$ (dashed curves).
In Fig. 2b, the $95\%$ C.L.
allowed region for the anomalous electric dipole moment
lies beneath the curves.  The bounds on the chromo-dipole moments are found
to be weak, since they only enter the decay rate via operator mixing.
For $m_t=150\gev$, $\kappa_\gamma$ is constrained to lie in the range
$(-2.6~{\rm to}~3.4)\times 10^{-16}$ e-cm, and $\tilde\kappa_\gamma<5.1
\times 10^{-16}$ e-cm.

The chiral structure of the top-bottom charged current is also probed by \bsg.
It has been determined\cite{fuji} that
consistency with the CLEO results restricts the potential deviation from the
$v-a$ structure of the $tbW$ coupling to be less than a few percent.

\subsection{Anomalous Trilinear Gauge Couplings}
The trilinear gauge coupling of the photon to $W^+W^-$ can also be tested
by the \bsg\ process.  Anomalous $\gamma WW$ vertices can be probed by
looking for deviations from the SM in tree-level processes such as
$\epem\to W^+W^-$ and $p\bar p \to W\gamma$, or by their influence on
loop order processes, for example the $g-2$ of the muon.  In the latter case,
cutoffs must be used in order to regulate the divergent loop integrals
and can introduce errors by attributing a physical significance to the
cutoff\cite{burgess}.  However, some loop processes, such as \bsg,
avoid this problem due to cancellations provided by the GIM mechanism
and hence yield cutoff independent bounds on anomalous couplings.
The CP-conserving interaction Lagrangian for $WW\gamma$ interactions is
\begin{equation}
{\cal L}_{WW\gamma}= i\left( W^\dagger_{\mu\nu}W^\mu A^\nu-W^\dagger_\mu
A_\nu W^{\mu\nu}\right) +i\kappa_\gamma W^\dagger_\mu W_\nu A^{\mu\nu}
+i{\lambda_\gamma\over\mws}W^\dagger_{\lambda\mu}W^\mu_\nu A^{\nu\lambda}
+h.c. \,,
\end{equation}
where $V_{\mu\nu}=\partial_\mu V_\nu-\partial_\nu V_\mu$, and the two
parameters $\kappa_\gamma=1+\Delta\kappa_\gamma$ and $\lambda_\gamma$ take on
the values $\Delta\kappa_\gamma, \lambda_\gamma=0$ in the SM.  In this case,
only the coefficient
of the magnetic dipole $b\to s$ transition operator $O_7$ is modified
by the presence of these additional terms and can be written as
\begin{equation}
c_7(M_W) = G_7^{SM}(\mts/\mws) +\Delta\kappa_\gamma A_1(\mts/\mws)
+\lambda_\gamma A_2(\mts/\mws) \,.
\end{equation}
The functions $A_{1,2}$ are obtained in the same manner as described above for
the anomalous top-quark couplings and are given explicitly in
Ref. 17.  As both of these parameters are varied, either large
enhancements or suppressions over the SM prediction for the \bsg\ branching
fraction can be obtained.  When one demands consistency with both the upper
and lower CLEO bounds, a large region of the
$\Delta\kappa_\gamma-\lambda_\gamma$ parameter plane is excluded; this
is displayed in Fig. 3 from Rizzo\cite{tgr} for $m_t=150\gev$.
Here, the $95\%$ C.L. bounds
obtained from the lower limit on $B(\bsg)$ correspond to the dashed curves,
where the region between the curves is excluded, while the constraints placed
from the upper CLEO limit correspond to the diagonal solid lines, with the
allowed region lying in between the lines.  The allowed region in this
parameter plane as determined from UA2 data\cite{uatwo} from the reaction
$pp\to W\gamma$ is also displayed in this figure and corresponds to the
region between the two almost horizontal lines.  Combining these constraints,
an overall allowed region is obtained and is represented by the two
shaded areas in this figure.  We see that a sizable area of the parameter
space is ruled out!  Note that the SM point in the $\Delta\kappa_\gamma-
\lambda_\gamma$ plane (labeled by `S') lies in the center of one of the
allowed regions.

\subsection{Fourth Generation}

The implications of a fourth generation of quarks on the process \bsg\
have been previously\cite{bsgfour} examined.  The possibility of a fourth
family of fermions was a popular\cite{santamon} potential extension to the
SM before LEP/SLC data\cite{lep} precluded the existence of a light fourth
neutrino.  However, one should keep in mind that a fourth generation is
consistent with the LEP/SLC data as long as the fourth neutrino is heavy, \ie,
$m_{\nu_4}\gsim M_Z/2$, and that such a heavy fourth neutrino could
mediate\cite{seesaw} a see-saw type mechanism thus generating a small mass for
$\nu_{e, \mu, \tau}$.

In the case of four families there is an additional contribution to \bsg\ from
the virtual exchange of the fourth generation up quark $t'$.  The Wilson
coefficients of the dipole operators are given by
\begin{equation}
c_{7,8}(M_W)=G_{7,8}(m_t^2/M_W^2)+{V^*_{t's}V_{t'b}\over V^*_{ts}V_{tb}}
G_{7,8}(m_{t'}^2/M_W^2) \,,
\end{equation}
in the limit of vanishing up and charm quark masses.  $V_{ij}$ represents the
4x4 CKM matrix which now contains 9 parameters; 6 angles and 3 phases.  We
recall here that the CKM coefficient corresponding to the t-quark contribution,
\ie, $V^*_{ts}V_{tb}$, is factorized in the effective Hamiltonian as shown in
Eqn. (1).  In order to determine the allowed ranges of the nine parameters
in the full 4x4 CKM matrix we demand consistency\cite{pdg} with (i) unitarity
and the determination of the CKM matrix elements extracted from charged current
measurements, (ii) the ratio $|V_{ub}|/|V_{cb}|$, (iii)
$\epsilon$, (iv) $B^0-\bar B^0$ mixing. $10^8$ sets of the nine CKM mixing
parameters are generated via Monte Carlo and subjected to the constraints
(i)-(iv) for $m_t=130-200 \gev$ and $m_{t'}=200-400\gev$.
The surviving sets of CKM parameters are then used to calculate the range
of $B(\bsg)$ in the four generation standard model.  This branching fraction
is displayed in Fig. 4 as a function of $m_t$ where the vertical lines
represent the allowed fourth
generation range as $m_{t'}$ is varied in the above region, and the solid
curve corresponds to the three generation value.  We see that once the
restrictions (i)-(iv) above are applied, the four generation \bsg\
branching fraction is essentially (except for smaller values of $m_t$)
within the range allowed by CLEO.

\subsection{Two-Higgs-Doublet Models}

Next we turn to two-Higgs-doublet models (2HDM), where we examine two distinct
models which naturally avoid tree-level flavor changing neutral currents.
In Model I, one doublet ($\phi_2$) generates masses for all fermions and the
other doublet ($\phi_1$) decouples from the fermion sector.  In the second
model (Model II) $\phi_2$ gives mass to the up-type quarks, while the
down-type quarks and charged leptons receive their mass from $\phi_1$.  Each
doublet obtains a vacuum expectation value (vev) $v_i$, subject to the
constraint that $v_1^2+v_2^2=v^2$, where $v$ is the usual vev present in the
SM.  The charged Higgs boson interactions with
the quark sector are governed by the Lagrangian
\begin{equation}
{\cal L}= {g\over 2\sqrt 2M_W} \ch\left[ V_{ij}m_{u_i}A_u\bar u_i(1-\gamma_5)
d_j+V_{ij}m_{d_j}A_d\bar u_i(1+\gamma_5)d_j \right] + h.c. \,,
\end{equation}
where $g$ is the usual SU(2) coupling constant and $V_{ij}$ represents the
appropriate CKM element.  In model I, $A_u=\cot\beta$ and
$A_d=-\cot\beta$, while in model II, $A_u=\cot\beta$ and $A_d=\tan\beta$,
where $\tan\beta\equiv v_2/v_1$ is the ratio of vevs.   In
both models, the \ch\ contributes to \bsg\ via virtual
exchange together with the top-quark, and the dipole $b\to s$ operators
$O_{7,8}$ receive contributions from this exchange.  At the $W$ scale the
coefficients of these operators take the generic form
\begin{equation}
c_{7,8}(M_W)=G_{7,8}(\mts/\mws) +{1\over 3\tan^2\beta}G_{7,8}(\mts/\mchs)
+\lambda F_{7,8}(\mts/\mchs) \,,
\end{equation}
where $\lambda=-1/\tan\beta,~+1$ in Model I and II, respectively.
The analytic
form of the functions $F_{7,8}$ can be found in Ref.\ 23.
Since the \ch\ contributions all scale as $\cot^2\beta$ in Model I,
enhancements to the SM decay rate only occurs for small values of \tb.
The relative minus sign between the two \ch\ contributions in this model
also gives
a destructive interference for some values of the parameters.
Consistency with the
CLEO lower and upper limits excludes\cite{me}
the shaded regions in the $\mch-\tb$
parameter plane presented in Fig.
5a, assuming $m_t=150\gev$.  Here, the shaded region on the left results from
the CLEO upper bound and the shaded slice in the middle is from the lower
limit.
In Model II, large enhancements also appear for small values of \tb, but
more importantly, $B(\bsg)$ is always larger than that of the SM,
independent of the value of \tb.  This is due to the $+\tan\beta$ scaling of
the $F_{7,8}$ term in Eq. (9).  In this case the CLEO upper bound
excludes\cite{me,vb}
the region to the left and beneath the curves shown in Fig. 5b for the various
values of $m_t$ as indicated.  In this case, the bounds
are quite sensitive\cite{bsgqcd} to the uncertainties arising from the higher
order QCD corrections.  We note that the \ch\ couplings present in
Model II are of the type present in Supersymmetry.   However, the limits
obtained in supersymmetric theories also depend on the size of the other
super-particle contributions to \bsg\  and are generally much more
complex\cite{bert} as discussed below.

\subsection{Supersymmetry}

In the supersymmetric standard model flavor mixing is also present in the
squark sector and hence flavor changing neutral
current processes are sensitive to the masses and mixings
of the super-partners.  For example, $K^0-\bar K^0$ mixing has been shown
to place\cite{susymix} stringent constraints on the level of degeneracy for
the first two generations of squarks (if one assumes CKM-like mixing).  One
should also be reminded, of course, that magnetic moment transition
operators, including \bsg,
vanish in the exact supersymmetric limit\cite{susylim}.

There are five classes of contributions
to \bsg\ in supersymmetric theories; the virtual exchange of (i) the up-type
quarks and the $W$ boson in the SM, (ii) the up-type quarks and the $H^\pm$
of Model II above, (iii) the up-type squarks and charginos, $\tilde\chi^\pm_i$,
(iv) the down-type squarks and neutralinos, $\tilde\chi^0_i$,
and (v) the down-type squarks and gluinos, $\tilde g$.  As discussed above
the contributions from (i) and (ii) are large and interfere
constructively.  It has been shown\cite{bert,susybsg} that contributions (iv)
and (v) are usually small in the minimal supersymmetric model and are not
competitive with those induced by $W$ boson and $H^\pm$
exchange.  However, the chargino contributions (iii) can be large, and for some
range of the parameter space can
cancel the $H^\pm$ contributions to give a value of $B(\bsg)$ at or even below
the SM prediction.

Several recent analyses of the chargino contributions have appeared in the
literature\cite{susylim,susybsg}.
The size and relative sign of these contributions
depend on the parameters present in the chargino mass matrix and on those
responsible for the masses and mixings of the squark sector.  Assuming
unification at a high energy scale, we take these parameters to be
the common soft-breaking gaugino mass $m_\lambda$, the universal scalar mass
$m_0$, the supersymmetric higgsino mass parameter $\mu$, the universal
trilinear soft-breaking scalar term in the superpotential $A$, $\tan\beta$,
as well as $m_t$.  Here we will consider the case where the up and charm
squark masses are degenerate, and will examine the effects of the possibly
large
stop-squark mass splitting due
to the potentially sizeable off-diagonal terms in the stop mass matrix.
The chargino-squark contributions to the Wilson coefficients
for the $b\to s$ transition dipole operators are given by\cite{susylim,susybsg}
\begin{eqnarray}
c_{7,8}^{\tilde\chi^\pm}(M_W) & \simeq & \sum_{j=1}^2 \left\{
{M_W^2\over \tilde m^2_{\chi_j^\pm}} |V_{j1}|^2G_{7,8}
\left( {\tilde m^2\over \tilde m^2_{\chi_j^\pm}} \right)
-{M_WU_{j2}V_{j1}\over \tilde m_{\chi_j^\pm}\sqrt 2\cos\beta} H_{7,8}
\left( {\tilde m^2\over \tilde m^2_{\chi_j^\pm}} \right) \right. \nonumber \\
& & +\sum_{k=1}^2 \left[ -{M_W^2\over \tilde m^2_{\chi_j^\pm}}
\left|V_{j1}T_{k1}-{m_tV_{j2}T_{k2}\over M_W\sqrt 2\sin\beta}\right|^2G_{7,8}
\left( {\tilde m^2_{t_k}\over \tilde m^2_{\chi_j^\pm}} \right) \right. \\
& & \left.\left. + {M_WU_{j2}T_{k1}\over \tilde m_{\chi_j^\pm}\sqrt 2\cos\beta}
\left(V_{j1}T_{k1}-{m_tV_{j2}T_{k2}\over M_W\sqrt 2\sin\beta}\right)H_{7,8}
\left( {\tilde m^2\over \tilde m^2_{\chi_j^\pm}} \right)
\right] \right\} \,,\nonumber
\end{eqnarray}
where $\tilde m_{\chi_j^\pm}$ represents the chargino masses, $\tilde m$
the up and charm squark masses, $\tilde m_{t_k}$ the stop-squark
masses, $U_{ij}$ and $V_{ij}$ are the unitary matrices which diagonalize
the chargino mass matrix, and $T_{kl}$ diagonalizes the stop-squark mass
matrix.  These all are calculable in terms of the supersymmetry parameters
listed above.  The functions $H_{7,8}$ are given in Refs.\ 26,28,29.
Contours of $B(\bsg)$, including the SM, $H^\pm$, and $\tilde\chi^\pm$
contributions, are displayed in Fig. 6 from Garisto and
Ng\cite{susybsg} in the $m_\lambda - \mu$ parameter plane for four
values of $A=\pm 1, \pm 2$ and taking $m_0=100\gev,\ m_t=140\gev$,
and $\tan\beta=10$.  It is immediately clear from the figure, that
regions of parameter space do exist where $B(\bsg)_{\rm SUSY}$ is at or
below the SM value and is consistent with the CLEO bounds.  It is found that
the stop-squark and chargino contributions have a large destructive
interfere with the SM and $H^\pm$ contributions when $\tilde t_1$ is
light (\ie, when there is a large stop mass splitting), $\tan\beta$ is large,
and $A\mu<0$.  However, if all the up-type squarks are degenerate,
the chargino contributions exactly cancel due to a SUSY-GIM
mechanism.  In this case, the $H^\pm$ mass is constrained to be large as
shown in the previous section.

\subsection{Three-Higgs-Doublet Models}

New CP violating phases are present in models with three or more scalar
doublets.  These phases appear in charged scalar exchange
and can influence CP asymmetries in neutral $B$ decays, even if the Yukawa
couplings obey natural flavor conservation\cite{thhiggs}.  For example, in a
three-Higgs-Doublet model (3HDM) one can avoid tree-level flavor
changing neutral currents by requiring
that a different doublet generate a mass for the up-type quarks, the down-type
quarks, and the charged leptons, respectively.  In this case, the interaction
Lagrangian between the quark sector and the two physical charged Higgs bosons
is
written as\cite{abs}
\begin{equation}
{\cal L} = {g\over 2M_W} \sum_{i=1,2} H^+_i\bar U \left[ Y_iM_uV_{CKM}(1-
\gamma_5) + X_iM_dV_{CKM}(1+\gamma_5)\right] D + h.c. \,,
\end{equation}
where $X$ and $Y$ are complex coupling constants that arise from the
diagonalization of the charged scalar mixing matrix and obey the relation
\begin{equation}
\sum_{i=1,2} X_iY_i^* =1 \,.
\end{equation}
Both $H^\pm_1$ and $H^\pm_2$ contribute to \bsg\ and
the Wilson coefficients $c_{7,8}$ at the matching scale $M_W$ now become
\begin{equation}
c_{7,8}(M_W)=G_{7,8}(\mts/\mws)+ \sum_{i=1,2} \left[ {|Y_i|^2\over 3}
G_{7,8}(m_t^2/m^2_{H^\pm_i}) + X_iY_i^* F_{7,8}(m_t^2/m^2_{H^\pm_i})\right] \,,
\end{equation}
with the analytic expressions for the functions $F_{7,8}$ being the same as in
the two-Higgs-Doublet case\cite{bsgch}.  The $X_iY_i^*$ term signals the
existence of a relative phase in the \bsg\ amplitude.
When evolved down to the $b$-quark scale, the contributions proportional
to $\im(X_iY^*_i)$ do not interfere with the remaining terms in
$c_{7,8}(M_W)$ and do not mix with the 4-quark operators.  Hence
these terms only appear quadratically in the expression for the \bsg\ rate.
A conservative upper limit can be placed on the value of $|\im(XY^*)|$
(where $\im(XY^*)=\im(X_1Y_1^*)=-\im(X_2Y_2^*)$ as given in Eqn. (12) above) by
letting the imaginary contribution alone saturate the CLEO upper bound. These
constraints are displayed in Fig. 7 as a function of the lightest charged
Higgs mass $m_{H^\pm_1}$ for various values of the heavier charged
Higgs mass $m_{H^\pm_2}$, subject to the restraint $m_{H^\pm_1}<m_{H^\pm_2}$.
The bottom solid curve corresponds to the case where the contribution of
the second charged Higgs $H^\pm_2$ is neglected.  We see that the constraints
depend very strongly on the value of $m_{H^\pm_2}$ and that the bounds
disappear when $m_{H^\pm_1}\simeq m_{H^\pm_2}$ due to an exact cancellation
between the two $H^\pm_i$ contributions.

\subsection{Extended Technicolor}

The decay \bsg\ has been investigated within the framework of various classes
of Extended Technicolor (ETC) models in Ref. 32. These
contributions were found to be either comparable or suppressed relative
to those of the SM, since gauge invariance implies that the photon vertex
is corrected only at higher order in these models.  We note that the $Z$-boson
couplings are modified at leading order in these theories and that large rates
for the decays $B\to \mu\mu$ and $b\to s\mu\mu$ can be
obtained\cite{etc,etctwo}.  The effective Lagrangian for ETC gauge boson
exchange in these scenarios can be written as
\begin{equation}
{\cal L}={1\over f^2}(\bar\psi^i_L\gamma_\mu T_L)(\bar U_R\gamma^\mu u^j_R)
Y_u^{ij} + {1\over f^2}(\bar\psi^i_L\gamma_\mu T_L)(\bar D_R\gamma^\mu
d^j_R)Y_d^{ij} + h.c. \,,
\end{equation}
where $T_L$ is a techni-doublet with the right-handed techni-partners
$U_R$ and $D_R$, $\psi_L$ represents the left-handed quark doublets with
$u_R$ and $d_R$ being the right-handed partners, the matrices
$Y_{u,d}^{ij}$ parameterize the symmetry breaking, and $i,j$ are generation
indices.

The first class of models considered in Ref. 32 is that of
``traditional'' ETC, which contains the minimal set of interactions necessary
to generate the third generation quark masses.  In this case the ETC gauge
boson spectrum is highly non-degenerate and the quark mass matrices are
approximately given by
\begin{equation}
M_{u,d}\sim {4\pi v^3\over f^2} Y_{u,d}\,.
\end{equation}
Working in the basis where $Y_u^{33}$ is normalized to unity, gives the
relation
$f^2 \sim 4\pi v^3/m_t$.  The dominant contribution to \bsg\ occurs when the
ETC gauge boson is exchanged between purely left-handed doublets and when the
photon is emitted from the technifermion line.  This results in the magnetic
moment operator
\begin{equation}
{m_t\over 4\pi v} {m_bV_{ts}\over (4\pi v)^2} \bar b_R\sigma^{\mu\nu}
s_L {e\over 2}f_{\mu\nu} \,.
\end{equation}
Comparing this to the corresponding quantity in the SM (\ie, $c_7(M_W)O_7$)
shows\cite{etc} that the ETC contribution
is suppressed with respect to the SM by a factor of
$m_t/[4\pi vG_7(\mts/\mws)]$.

The second class of models considered in this reference are those which
incorporate a techni-GIM mechanism which provide a GIM-like suppression of
flavor changing
neutral currents due to a restricted form of flavor symmetry breaking.
The ETC scale becomes $f^2=m^2_{ETC}/g^2_{ETC}$, where $m_{ETC} (g_{ETC})$
represents the mass (coupling) of the nearly degenerate ETC gauge bosons.
Here, the dominant contribution to \bsg\ results when the photon is
attached to the ETC gauge boson line.  Assuming $Y_u^{23}\sim V_{ts}$,
the effective magnetic moment operator is estimated\cite{etc} to be
\begin{equation}
{\xi_4m_bV_{ts}\over 4m^2_{ETC}}\bar b_R\sigma_{\mu\nu}s_L
{e\over 6}F^{\mu\nu} \,,
\end{equation}
with $\xi_4$ being a model dependent parameter.  This contribution is
expected\cite{etc} to yield a rate for \bsg\ which is within $10\%$ of that in
the SM.

\subsection{Leptoquarks}

Leptoquarks are color triplet particles which couple to a lepton-quark pair
and are naturally present in many theories beyond the SM which relate leptons
and quarks at a more fundamental level.  They appear in technicolor theories,
models with quark-lepton substructure, horizontal symmetries, and grand
unified theories based on the gauge groups SU(5), SO(10), and $E_6$.  In all
these scenarios leptoquarks carry both baryon and lepton number but their
other quantum numbers, \ie, spin, weak isospin, and electric charge, vary
between the different models.   The scalar and vector leptoquark interaction
Lagrangians\cite{buch,sacha} which are renormalizable, baryon and lepton
number conserving, and consistent with the symmetries of
$SU(3)_C\times SU(2)_L\times U(1)_Y$ are given by
\begin{eqnarray}
{\cal L}_S & = & (\lambda_{LS_0}\bar q^c_Li\tau_2\ell_L+
\lambda_{RS_0}
\bar u^c_Re_R)S^\dagger_0 + \lambda_{R\tilde S_0}\bar
d^c_Re_R\tilde S_0^\dagger
+(\lambda_{LS_{1/2}}\bar u_R\ell_L \\
& & +\lambda_{RS_{1/2}}\bar q_Li\tau_2e_R)S^\dagger_{1/2} +
\lambda_{L\tilde S_{1/2}}\bar d_R\ell_L\tilde S^\dagger_{1/2} +
\lambda_{LS_1}\bar q^c_Li\tau_2\vec\tau\ell_L\cdot\vec S_1^\dagger
+ h.c. \,, \nonumber \\
{\cal L}_V & = & (\lambda_{LV_0}\bar q_L\gamma_\mu\ell_L + \lambda_{RV_0}
\bar d_R\gamma_\mu e_R)V_0^{\mu\dagger} + \lambda_{R\tilde V_0}\bar u_R
\gamma_\mu e_R\tilde V_0^{\mu\dagger} + (\lambda_{LV_{1/2}}\bar d^c_R\gamma_\mu
\ell_L \nonumber \\
& & + \lambda_{RV_{1/2}}\bar q^c_L\gamma_\mu e_R)V_{1/2}^{\mu\dagger}
+\lambda_{L\tilde V_{1/2}}\bar u^c_R\gamma_\mu\ell_L\tilde V_{1/2}^{\mu\dagger}
+\lambda_{LV_1}\bar q_L\gamma_\mu\vec\tau\ell_L\cdot\vec V_1^{\mu\dagger}
+ h.c. \,. \nonumber
\end{eqnarray}
Here the subscripts 0, 1/2, and 1 represent the SU(2) singlet, doublet, and
triplet leptoquarks, respectively, the $\lambda$'s are {\it a priori} unknown
Yukawa coupling constants, the $L(R)$ index on the coupling reflects
the chirality of the lepton, and the generation indices have been suppressed.

Leptoquarks can contribute to \bsg\ by the virtual exchange of a charged lepton
and a leptoquark in a penguin diagram.  These diagrams have been calculated
in Ref.\ 35, where the leptoquark contributions to $c_8(M_W)$ have
been neglected.  Using the approximation that the leptoquark contributions to
the \bsg\ amplitude must be smaller than that for the SM,
Davidson \etal\cite{sacha} derive the following bounds on the relevant
combinations of the Yukawa coupling constants for scalar leptoquarks,
\begin{equation}
\lambda_L^{\ell b}\lambda_L^{\ell s}, \lambda_R^{\ell b}\lambda_R^{\ell s}
< {3\times 10^{-2}\over Q_\ell +{1\over 2}Q_{LQ}} \left( {m_{LQ}\over
100\gev}\right)^2 \,,
\end{equation}
where $\ell$ is a charged lepton of any generation, and $Q_\ell(Q_{LQ})$
are the electric charges of the exchanged lepton (leptoquark).
Similarly,  for non-gauge vector leptoquarks,
\begin{equation}
\lambda_L^{\ell b}\lambda_L^{\ell s}, \lambda_R^{\ell b}\lambda_R^{\ell s}
< {2\times 10^{-2}\over 2Q_\ell +{5\over 2}Q_{LQ}} \left( {m_{LQ}\over
100\gev}\right)^2 \,.
\end{equation}
We note that other $B$ decays, such as $B\to \ell^-\ell^{(')+}$, can
provide stronger constraints\cite{sacha} on these leptoquark couplings.

\subsection{Left-Right Symmetric Models}

The last scenario of new physics that we will consider is the Left-Right
Symmetric Model (LRM)\cite{lrm} which is based on the extended gauge
group $SU(2)_L\times SU(2)_R\times U(1)$.  Such theories have been popular
for many years, as both a possible generalization of the SM and in the
context of grand unified theories such as $SO(10)$ and $E_6$.  One
prediction of these models is the existence of a heavy, charged,
right-handed gauge boson $W_R^\pm$, which in principal mixes with the
SM $W_L^\pm$ via a mixing angle $\phi$ to form the mass eigenstates
$W_{1,2}^\pm$.  This mixing angle is constrained\cite{phidata} by data in
polarized $\mu$ decay (in the case of light right-handed neutrinos) and from
universality requirements to be $|\phi|\lsim 0.05$.
The exchange of a $W_R^\pm$ within a penguin diagram,
in analogy with the SM $W_L^\pm$ exchange, can lead to significant
deviations from the SM prediction for the rate in \bsg\ which are
sensitive to the sign and magnitude of the angle $\phi$.

The first class of LRM we will discuss is that where the
right-handed and left-handed CKM mixing matrices are assumed to be equal,
\ie, $V_R=V_L$.  In this case, $W_R^\pm$ searches at the Tevatron collider
together with the value of the $K_L-K_S$ mass difference
constrain\cite{wrdata} the mass of $W_R^\pm$ to be at least $m_{W_R}>
1.6\kappa\tev$, where $\kappa\equiv g_R/g_L$ is the ratio of
right-handed to left-handed $SU(2)$ coupling constants.  In the LRM the
complete operator basis governing $b\to s$ transitions is expanded to include
20 operators.  Two new four-quark operators $O_{9,10}$ which have different
chirality structure are also present,\cite{tgrlrm,lrmbsg} and  left-right
symmetry dictates the existence of a set of operators which have a flipped
chirality structure compared to the standard set.  The latter are obtained
by the substitution $P_L\leftrightarrow P_R$ in the definition of
$O_{1-10}$, where $P_{L,R}=(1\pm\gamma_5)/2$.
We denote the standard set of operators as ``left-handed'', \ie, $O_{1L-10L}$,
and the chirality flipped operators as ``right-handed'', $O_{1R-10R}$.
These two sets of operators do not mix under the QCD evolution and thus
can be treated independently.  The expression for the partial decay
width now becomes
\begin{equation}
\Gamma(\bsg)={\alpha(m_b)G_Fm_b^5\over 128\pi^4}|V_L^{ts*}V_L^{tb}|^2
\left( |c_{7L}(m_b)|^2 + |c_{7R}(m_b)|^2\right) \,,
\end{equation}
where $c_{7L,R}(M_{W_1})$ are defined via the low-energy effective
Hamiltonian
\begin{equation}
{\cal H}_{eff} = -{G_Fem_b\over 4\sqrt 2\pi^2}\bar s\sigma_{\mu\nu}
\left( c_{7L}P_R + c_{7R}P_L \right) bF^{\mu\nu} \,.
\end{equation}
The expressions for the coefficients $c_{iL}$ and $c_{iR}$ are evaluated via
the one-loop matching conditions at the scale $M_{W_1}$ and are given in
Ref. 39.  The branching fraction is obtained by scaling to
the semi-leptonic decay rate as usual, except that possible $W_R^\pm$
contributions\cite{lrmsemi} to $b\to c\ell\nu$ must also be included.
The resulting values for $B(\bsg)$ from the work of
Rizzo\cite{tgrlrm}  are displayed in Fig. 8a-b as a function of the tangent
of the $W_L-W_R$ mixing angle, $\tan\phi$, for $M_{W_R}=1.6\tev$.
Figure 8a examines the branching fraction for various values of the
top-quark mass  assuming $\kappa=1$, while Fig.\ 8b fixes $m_t=160\gev$
and varies $\kappa$ between 0.6 and 2.  In both cases the solid
horizontal line represents the CLEO bound.  We see from the figures that
for $\kappa=1$, $\tan\phi$ is constrained to lie in the range
$-0.02\lsim\tan\phi\lsim 0.005$ and that these bounds strengthen with
increasing values of $\kappa$.  We also note that \bsg\ was found not
to be sensitive to the exact value of the $W_R$ mass.

The assumption that $V_R=V_L$ is simple and attractive, but one should
keep in mind that realistic and phenomenologically viable models can be
constructed where $V_R$ is unrelated to $V_L$, and hence the above constraints
on the model parameters can be avoided.  One such example is the
model given in Gronau and Wakaizumi\cite{gron}, where $B$ decays proceed
only through right-handed currents.  Using the form of $V_L$ and $V_R$
given in this reference, consistency with the $B$ lifetime provides the
bound $M_{W_R}\leq 416.2\kappa [|V_R^{cb}|/\sqrt 2]^{1/2}\simeq 415
\kappa\gev$.  Collider bounds from the Tevatron can be satisfied\cite{wrdata}
in this model if $\kappa\geq 1.5$ and $M_R\geq 600\gev$, assuming that the
$W_R^\pm$ decays only into the known SM fermions as well as the right-handed
neutrino.  The value of $B(\bsg)$ as a function of $\tan\phi$ is presented
in Fig.\ 8c from Ref.\ 39.  Here, $\kappa=1.5$ and $M_{W_R}=
600\gev$ is assumed and the outer(inner)-most solid line corresponds to
$m_t=120(200)\gev$.  It is immediately clear that the allowed range of
$\tan\phi$ is much more restricted than in the $V_L=V_R$ case and that both
the upper and lower CLEO bounds will play a role.  For
$M_{W_R}=600(800)\gev$, the allowed ranges of $\tan\phi$ are found to be
$-0.43\times 10^{-3}<\tan\phi<0$, and $0.40\times 10^{-3}<\tan\phi<0.81\times
10^{-3}$ ($-0.32\times 10^{-3}<\tan\phi<0$, and $0.29\times 10^{-3}<\tan\phi<
0.60\times 10^{-3}$).  These ranges are highly constrained and a more precise
determination of the \bsg\ branching fraction would finely-tune the values
of the parameters in this model.

\section{Conclusion}

In summary, we have seen that the process \bsg\ provides powerful constraints
for a variety of models containing physics beyond the SM.
In most cases, these constraints
either complement or are stronger than those from other
low-energy processes and from direct collider searches.
We look forward to an exciting future in $B$ physics!

\def\MPL #1 #2 #3 {Mod.~Phys.~Lett.~{\bf#1},\ #2 (#3)}
\def\NPB #1 #2 #3 {Nucl.~Phys.~{\bf#1},\ #2 (#3)}
\def\PLB #1 #2 #3 {Phys.~Lett.~{\bf#1},\ #2 (#3)}
\def\PR #1 #2 #3 {Phys.~Rep.~{\bf#1},\ #2 (#3)}
\def\PRD #1 #2 #3 {Phys.~Rev.~{\bf#1},\ #2 (#3)}
\def\PRL #1 #2 #3 {Phys.~Rev.~Lett.~{\bf#1},\ #2 (#3)}
\def\RMP #1 #2 #3 {Rev.~Mod.~Phys.~{\bf#1},\ #2 (#3)}
\def\ZP #1 #2 #3 {Z.~Phys.~{\bf#1},\ #2 (#3)}

\newpage

%
\noindent
{\tenrm {Fig. 1.  The branching fraction for \bsg\ in the Standard Model
(a) as a function of the top-quark mass including QCD corrections to the
leading log (dashed) and next-to-leading log order (solid).  (b) Dependency of
the branching fraction on the choice of renormalization scale $\mu$ for various
values of the b-quark mass as indicated with $m_t=150\gev$.
}}
\vglue 0.2cm
\noindent
{\tenrm {Fig. 2.  The allowed range of (a) $\kappa_\gamma$ and (b) $
\tilde\kappa_\gamma$ assuming $^(\tilde\kappa^)_g=0$ (solid curve) or
$^(\tilde\kappa^)_g=^(\tilde\kappa^)_\gamma$ (dashed curve).
}}
\vglue 0.2cm
\noindent
{\tenrm {Fig. 3.  Allowed (shaded) region of the
$\Delta\kappa_\gamma-\lambda_\gamma$ parameter plane from the CLEO upper
and lower bounds on \bsg, assuming $m_t=150\gev$,
 and the UA2 event rate for $pp\to W\gamma$ as discussed in the text.
The point in this plane representing the SM is labeled by S.
}}
\vglue 0.2cm
\noindent
{\tenrm {Fig. 4.  Branching fraction for \bsg\ as a function of $m_t$.
The solid curve represents the three generation SM value and the vertical
lines are the allowed ranges of $B(\bsg)$ in the four generation model.}
\vglue 0.2cm
\noindent
{\tenrm {Fig. 5.  The excluded regions in the $m_{H^\pm}-\tan\beta$ plane
resulting from the present CLEO bounds in (a) Model\ I (shaded area is
excluded) for $m_t=150\gev$ and
(b) Model\ II for various values of $m_t$ as indicated, where the
excluded regions lie to the left and below each curve.}}
\vglue 0.2cm
\noindent
{\tenrm {Fig. 6.  Contours of $B(\bsg)$ in units of $10^{-4}$ in the
$m_\lambda - \mu$ parameter plane with $\tan\beta=10$, $m_t=140\gev$,
$m_0=100\gev$, taking $A=+1, -1 +2, -2$ in (a), (b), (c), (d), respectively.
All masses in GeV.}
\vglue 0.2cm
\noindent
{\tenrm {Fig. 7.  Constraints on $|\im(XY^*)|$ as a function of the mass
of the lightest charged Higgs boson, $m_{H^\pm_1}$ with $m_{H^\pm_2}=100,
250, 500, 750,\ {\rm and}\ 1000\gev$ corresponding (from left to right) to the
dashed, dash-dotted, solid, dotted, and dashed curves.  The bottom solid curve
represents the case where the $H^\pm_2$ contributions have been neglected.
The allowed region lies beneath the curves.}
\vglue 0.2cm
\noindent
{\tenrm {Fig. 8.  $B(\bsg)$ in the LRM assuming $V_R=V_L$ as a function
of the tangent of the $W_L-W_R$ mixing angle, $t_\phi$.  (a) $\kappa=1$ and
$M_{W_R}=1.6\tev$,
with $m_t=120, 140, 160, 180,$ and 200 GeV corresponding to the dotted,
dashed, dash-dotted, solid, and square-dotted curves, respectively.
(b) $m_t=160\gev$ and $M_{W_R}=1.6\tev$ with $\kappa$ varying between
0.6 (left-most dotted curve) and 2.0 (inner-most dash-dotted curve).
(c) Gronau-Wakaizumi version of the LRM with $\kappa=1.5$, $M_{W_R}=600\gev$
with the outer-most solid curve corresponding to $m_t=120\gev$, and is
increased in each case by steps of 20 GeV.}

\end{document}